\documentclass[aps, prb, preprint, superscriptaddress, nofootinbib, nobibnotes]{revtex4-1}

\usepackage{epsfig}
\usepackage{amsmath}
\usepackage{amssymb}
\usepackage{xcolor}
\usepackage{xspace}
\usepackage[sort&compress]{natbib}
\usepackage{cancel}
\usepackage[normalem]{ulem}
\usepackage{mathrsfs} 

\usepackage{times}
\usepackage[final]{pdfpages}
\usepackage{graphicx}
\usepackage{framed, color} 
\usepackage{capt-of}
\definecolor{shadecolor}{rgb}{0.875, 0.875, 0.875}

\usepackage{lineno}

\usepackage{hyperref}

\newcommand{\Eqref}[1]{\mbox{Eq.\hspace{0.25em}\eqref{#1}}}
\newcommand{\Eqsref}[1]{\mbox{Eqs.\hspace{0.25em}\eqref{#1}}}
\newcommand{\figref}[1]{\mbox{Fig.\hspace{0.25em}\ref{#1}}}
\newcommand{\vect}{\boldsymbol}
\newcommand{\Nabla}{\vect\nabla}
\newcommand{\eps}{\epsilon}
\newcommand{\diff}{\text{d}}

\newcommand{\nocontentsline}[3]{}
\newcommand{\tocless}[2]{\bgroup\let\addcontentsline=\nocontentsline#1{#2}\egroup}

\makeatletter
\def\l@subsubsection#1#2{}
\makeatother

\begin{document}

\title{Growth and Division of Active Droplets: A Model for Protocells} 
\date{\today}

\author{David Zwicker}
\thanks{These two authors contributed equally}
\affiliation{%
Max Planck Institute for the Physics of Complex Systems,
01187 Dresden, Germany}
\affiliation{%
School of Engineering and Applied Sciences, Harvard University, Cambridge, MA 02138, USA}

\author{Rabea Seyboldt}
\thanks{These two authors contributed equally}
\affiliation{%
Max Planck Institute for the Physics of Complex Systems,
01187 Dresden, Germany}

\author{Christoph A. Weber}
\affiliation{%
Max Planck Institute for the Physics of Complex Systems,
01187 Dresden, Germany}

\author{Anthony A. Hyman}
\affiliation{%
Max Planck Institute of Molecular Cell Biology and Genetics,
01307 Dresden, Germany}

\author{Frank J\"ulicher}
\email[To whom correspondence should be addressed; Email: ]{julicher@pks.mpg.de}
\affiliation{%
Max Planck Institute for the Physics of Complex Systems,
01187 Dresden, Germany}

\begin{abstract}
It has been proposed that during the early steps in the origin of life, 
small droplets could have formed 
via the segregation of molecules from complex mixtures by phase
separation. These droplets
could have provided
chemical reaction centers.
However,  whether 
these droplets could divide and propagate is unclear.  Here we examine the behavior of droplets in systems that are maintained away from thermodynamic equilibrium by an 
external supply of energy.  In these systems, droplets grow by the addition of
droplet material generated by chemical reactions. Surprisingly, we find that chemically driven droplet growth  
can lead to shape instabilities that trigger the division of droplets into two smaller daughters.  
Therefore, chemically active droplets can exhibit cycles of growth and division that resemble the proliferation of living cells. Dividing active droplets could serve as a model for prebiotic protocells, where chemical reactions in the droplet play the role of a prebiotic metabolism.\end{abstract}

\maketitle

\paragraph*{Introduction}
Living systems consist of cells that can grow and divide.
Cells take up matter from the outside world to grow, they release waste products, and they are able to divide, creating more cells. 
A fundamental question is to understand how cells arose early in evolution.
Early in the origin of life, chemical reaction centers or chemical micro reactors had to form in order to organize chemical reactions in space.
These micro reactors had to exchange material with the outside and they had to propagate.  
Recently, the idea of  
Oparin and Haldane~\cite{Oparin1952,haldane1929origin}
that small droplets, which they called coacervates, could organize molecules in micro reactors
has resurfaced to prominence\cite{Brangwynne2009,Koga2011,Crosby2012,Sokolova2013,Hyman2014,DoraTang2015}. 
Such droplets are liquid-like aggregates that concentrate molecules that have separated from a complex mixture.

Liquid droplets are self-organized structures that coexist with a surrounding fluid~\cite{Bray1994,Hyman2014}.
The interface separating the two coexisting phases provides them
 with a well defined surface. 
 The associated surface tension forces them into a spherical shape. Furthermore,
 many substances can diffuse across the interface. 
The segregation of components into a droplet concentrates material in a confined volume, which may facilitate specific chemical reactions.
Thus droplets provide containers in which chemical reactions can be spatially organized.
Although the thermodynamics of phase transitions can explain how liquid drops can form, it is unclear how such 
droplets could 
propagate by division and subsequent growth, an ability that would be key at the
origin of life. 

Droplets grow by taking up material from a supersaturated environment or
by Ostwald ripening~\cite{Ostwald1897,Lifshitz1961,Binder1976a,Voorhees1992,Bray1994}.
Ostwald ripening describes the exchange of material between droplets by diffusion, 
usually leading to growth of large droplets while small droplets shrink.
Furthermore, droplets can increase in size by fusion of two droplets into a larger one.
These processes lead to the formation of droplets of increasing size while the droplet number decreases with time.
This behavior is opposite to that of cells which have a characteristic size and increase their number by division.  How could droplets divide and propagate?

We have recently shown that droplets that are maintained away from thermodynamic equilibrium by a chemical fuel can have unusual  properties~\cite{Zwicker2014,Zwicker2015}.
In particular, in the presence of chemical reactions, Ostwald ripening can be suppressed~\cite{Zwicker2015} and multiple droplets can stably coexist, with a characteristic size set by the reaction rates~\cite{Puri1994,Glotzer1994,Carati1997,Zwicker2015}. 
Here, we show that surprisingly, spherical droplets subject to 
chemical reactions  spontaneously split in two smaller daughter droplets of equal size.
Therefore, chemically active droplets can grow and subsequently divide and thereby propagate by using up the inflowing material as a fuel. We conclude that droplets can indeed behave similarly to cells in the  presence of chemical reactions that are driven by an external fuel reservoir.  Such active droplets could represent models for growing and dividing protocells with a rudimentary metabolism 
which is represented by simple chemical reactions that are maintained by an external fuel.

\paragraph*{Division of active droplets}

Droplets can serve as small compartments to spatially organize chemical reactions. 
The emergence of droplets requires phase separation into two coexisting liquid phases 
of different composition.
Phase separation is driven by molecular interactions, where
molecules with an affinity for each other lower their energy if they come closely together.
A fluid can demix if the energy decrease associated with molecular interactions
overcomes the effects of entropy increase by mixing~\cite{Huggins1941,Flory1942}. 
If those interactions are strong, a sharp interface separates the coexisting phases.

\begin{figure}
	\centering
	\includegraphics[width=\hsize]{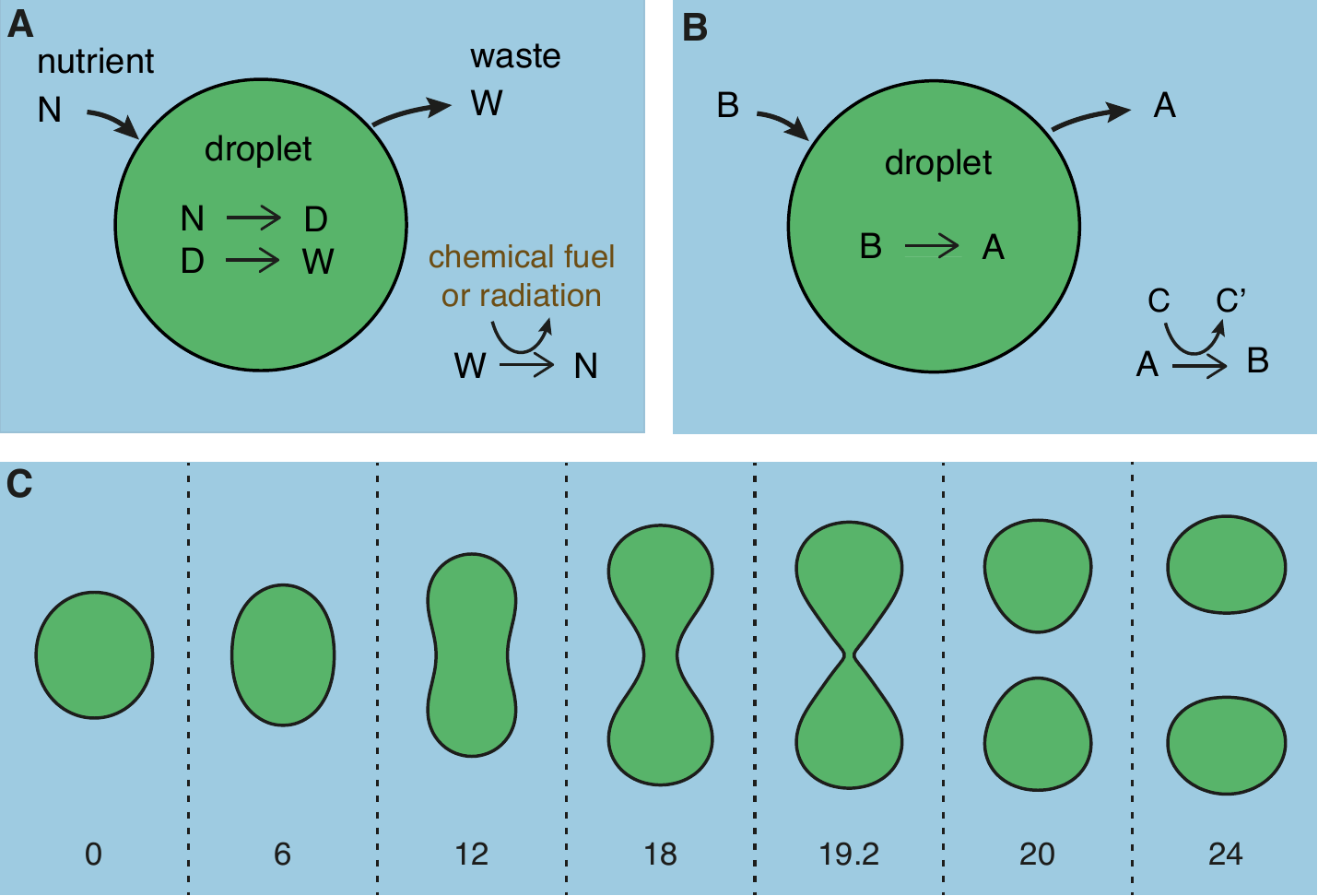}
	\caption[Protocells can divide in a minimal system]{
	A) Schematic representation of an active droplet as a simple model of a protocell.
	The droplet (orange) consists of a droplet material $D$. Nutrients $N$ of high chemical energy 
	can diffuse into the droplet. Inside the droplet,
	$N$ is transformed to $D$ by chemical reactions. Droplet material
	$D$ is degraded chemically into low energy waste $W$ 
	that leaves the droplet. 
	B) Minimal model, with droplet material $B$ and soluble component $A$.	
	C) Sequence of shapes of a dividing droplet at different times as indicated.
	The dynamic equations of a continuum model 
	corresponding to the situation shown in B) were solved numerically. 
	The droplet shapes are shown as equal concentration contours (black).
	Parameter values are $\nu_- t_0 / \Delta c  = 7 \cdot 10^{-3}$, 
	$\nu_+  t_0 /\Delta c = 1.9 \cdot 10^{-3} $, and $k_\pm t_0 = 10^{-2}$,
	where $t_0$ is a characteristic time of the continuum model  (see supplementary information). 
	Indicated times are given in units of $10^2 \, t_0$.
	}
	\label{fig1} \label{fig:schematic} \label{fig:snapshots}
\end{figure}

Droplets can become chemically active if the material of the droplet is produced and destroyed by chemical reactions.
An example that resembles a simple protocell is shown schematically in \figref{fig:schematic}A.
The droplet is  formed by a droplet material $D$ that 
 is generated inside the droplet 
from a high energy precursor $N$, which plays the role of a nutrient. 
Droplet material can degrade into a lower energy component $W$ 
that plays the role of a
waste, which leaves the droplet by diffusion. 
The droplet can survive if $N$ is continuously supplied and $W$ is continuously removed. This can be achieved by recycling $N$ using an external energy
source such as a fuel or radiation.

Inspired by Oparin \cite{Oparin1924a}, 
we discuss the physics of such active droplets using a minimal 
 model with only two components $A$ and  $B$, 
see \figref{fig:schematic}B.  The droplet material $B$ phase separates from the solvent.
It can spontaneously be degraded by a chemical reaction 
\begin{equation}
	B \rightarrow A
	\;
	\label{r1}
\end{equation}
into molecules of type $A$ that are soluble in the background fluid and leave the droplet. 
The backward reaction $A\rightarrow B$ is not proceeding spontaneously because
$B$ is of higher energy than $A$. New droplet material $B$ can be
produced by the second reaction
\begin{equation}
	A+C \rightarrow B+C' 
	 \; ,
	\label{r2}
\end{equation}
that is coupled to a fuel $C$.
Here $C'$ is the low energy reaction product of the fuel molecules. 
The chemical potential difference $\Delta\mu_C=\mu_C-\mu_{C'}>0$ provided by the fuel powers the production of
high energy $B$ from low energy $A$. 
The difference  
$\Delta\mu_C$ can be maintained constant if the concentrations of
$C$ and $C'$ are set by an external reservoir. 
In this case, the system is kept away from a thermodynamic equilibrium, see Box 1. 

The combination of phase separation and non-equilibrium chemical reactions can be 
studied in a continuum model \cite{Puri1994,Glotzer1994,Zwicker2015}, 
see supplemental information. 
Using this model, we find that spherical droplets that are chemically active can
undergo a shape instability and split in two smaller droplets, despite their surface tension,
see \figref{fig:schematic}C. 
A droplet first grows until it reaches its stationary size \cite{Zwicker2015}.
Then, the droplet starts to elongate and forms a dumbbell shape. 
This dumbbell splits in two smaller droplets of equal size. 
The resulting smaller droplets grow again until a new division may occur,
reminiscent of living cells.

In order to investigate the stability of spherical droplets, 
we study the droplet shape by an 
effective droplet model described in Box 2 \cite{Zwicker2015}.
\begin{figure*}
	\centering
	\includegraphics[width= \hsize]{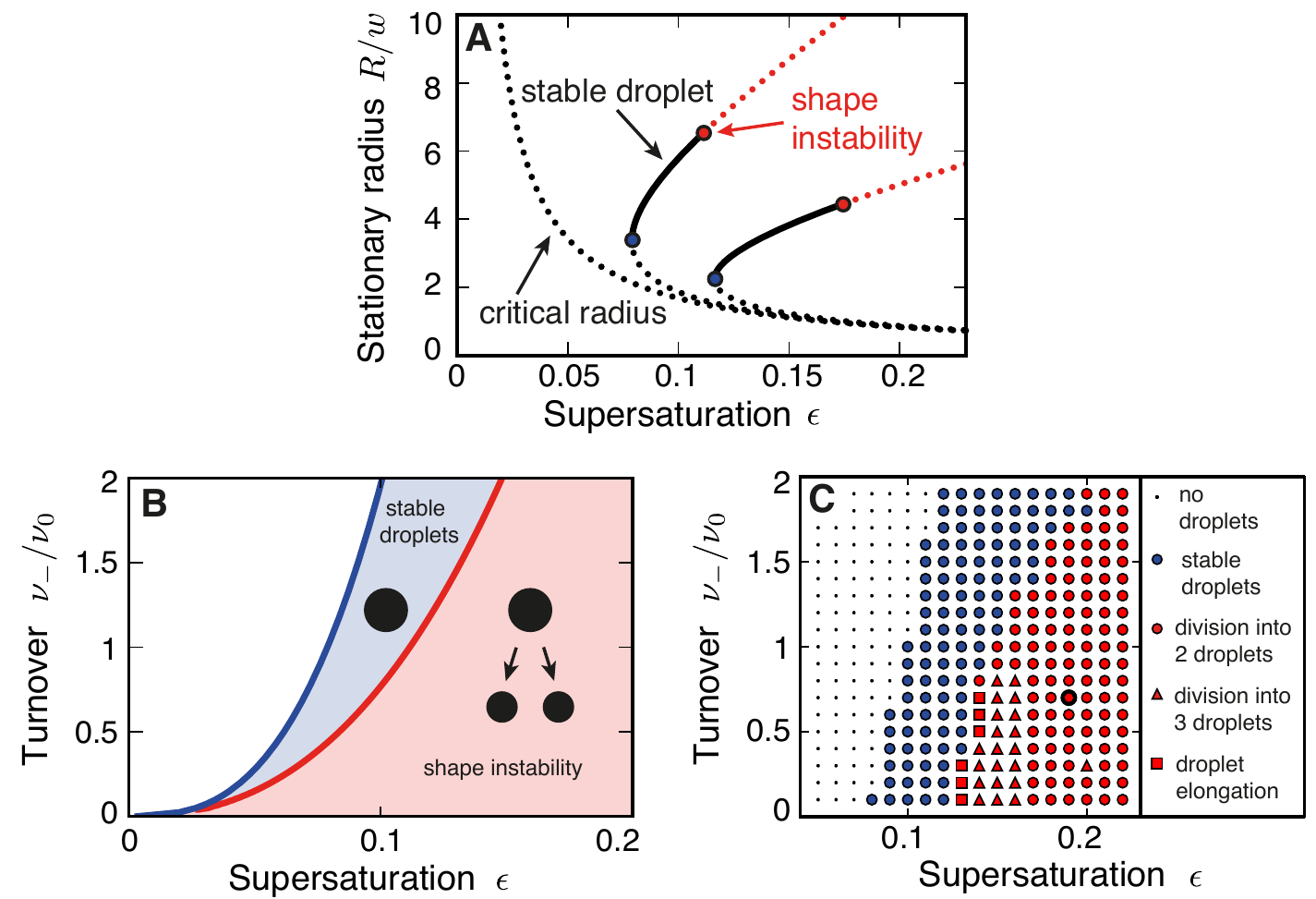}
	\caption[Stability diagram]{
	A) Stationary radii of active droplets. 
	The droplet radius $R$ of spherical droplets is shown as a function of 
	supersaturation $\eps$
	for different values of normalized turnover $\nu_-/\nu_0 = 0, 1, 3$ (from left to right). 
	Radii of stable droplets are shown as solid black lines.
	Dotted lines indicate states where droplets are unstable with respect to size (black) or shape (red).
	The results are obtained for the effective droplet model described in Box 2. 
         Parameter values are: $k_{\pm} \tau_0 = 10^{-2}$, $c_+^{(0)} =0$, 
         $\beta_-=\beta_+$, $D_-=D_+$ and $\nu_0 = 10^{-2}\Delta c/\tau_0$.
         Here, $w = 6 \beta_+ \gamma / \Delta c$, and $\tau_0 = w^2/D_+$ are
         characteristic length and time scales. 
         B) Stability diagram of active droplets as a function of 
         supersaturation $\eps=\nu_+/(k_+\Delta c)$ and
         turnover $\nu_-$ of droplet material.
         Droplets either dissolve and disappear (white region), are spherical and stable (blue region), or undergo a shape instability and typically divide (red region).
         The lines of instability are obtained for the droplet model described in 
         Box 2 for the same parameters as in A).
         C) Same stability diagram as in B) but for the continuum model described in the supplemental information.
         The behavior of droplets is indicated by symbols 
         for different values of $\nu_-$ and $\epsilon$. Parameter values are $k_\pm t_0 = 10^{-2}$
	(see supplementary information).   The parameter values corresponding to \figref{fig:snapshots}C are indicated 
         (large red circle).
         }
         \label{fig:stability_diagram}
\end{figure*}
\figref{fig:stability_diagram}A shows the behavior of the stationary 
droplet radius in this model as a function of the supersaturation  
$\eps$. This supersaturation is the excess concentration of droplet material far from the
droplet, generated by the chemical reaction~\eqref{r2}. 
For $\eps>0$, material diffuses to the droplet and is incorporated. 
\figref{fig:stability_diagram}A shows that for a given turnover $\nu_-$ of droplet material inside the droplet
(see Box 2),
stationary droplets only exist for sufficiently large supersaturation.
Beyond this threshold, droplets smaller than the
critical radius (\figref{fig:stability_diagram}A, black dotted lines) shrink, while
larger droplets grow toward the stationary radius 
(\figref{fig:stability_diagram}A, black solid line) \cite{Zwicker2015}.
At this stationary radius, the influx of $B$ due to supersaturation outside is 
balanced by the efflux of material $A$ produced inside the droplet. 
Thus a larger turnover leads to smaller droplets (\figref{fig:stability_diagram}A).

Droplet division occurs when a spherical droplet becomes unstable and elongates.
We performed a linear stability analysis
of spherical droplets
at their stationary radius in the effective droplet model, see supplemental material.
We find that for increasing supersaturation $\eps$, a spherical 
droplet with surface tension
undergoes a shape instability when its radius reaches a critical 
value $R_{\rm div}$ that depends
on the reaction rates and droplet parameters, see \figref{fig:stability_diagram}A.
Beyond the radius $R_{\rm div}$, the spherical shape is unstable and
any small shape deformation triggers the elongation of the droplet shape along one axis.

The stability analysis of the effective droplet model can be represented 
in a state diagram, see \figref{fig:stability_diagram}B. 
We find three different regions 
as a function of supersaturation $\eps$ and turnover of droplet material $\nu_-$.
A region where droplets do not exist (white), a region in which spherical droplets are stable (blue),
and a region in which spherical droplets are unstable (red).
\begin{figure}
	\centering
	\includegraphics[width= \hsize]{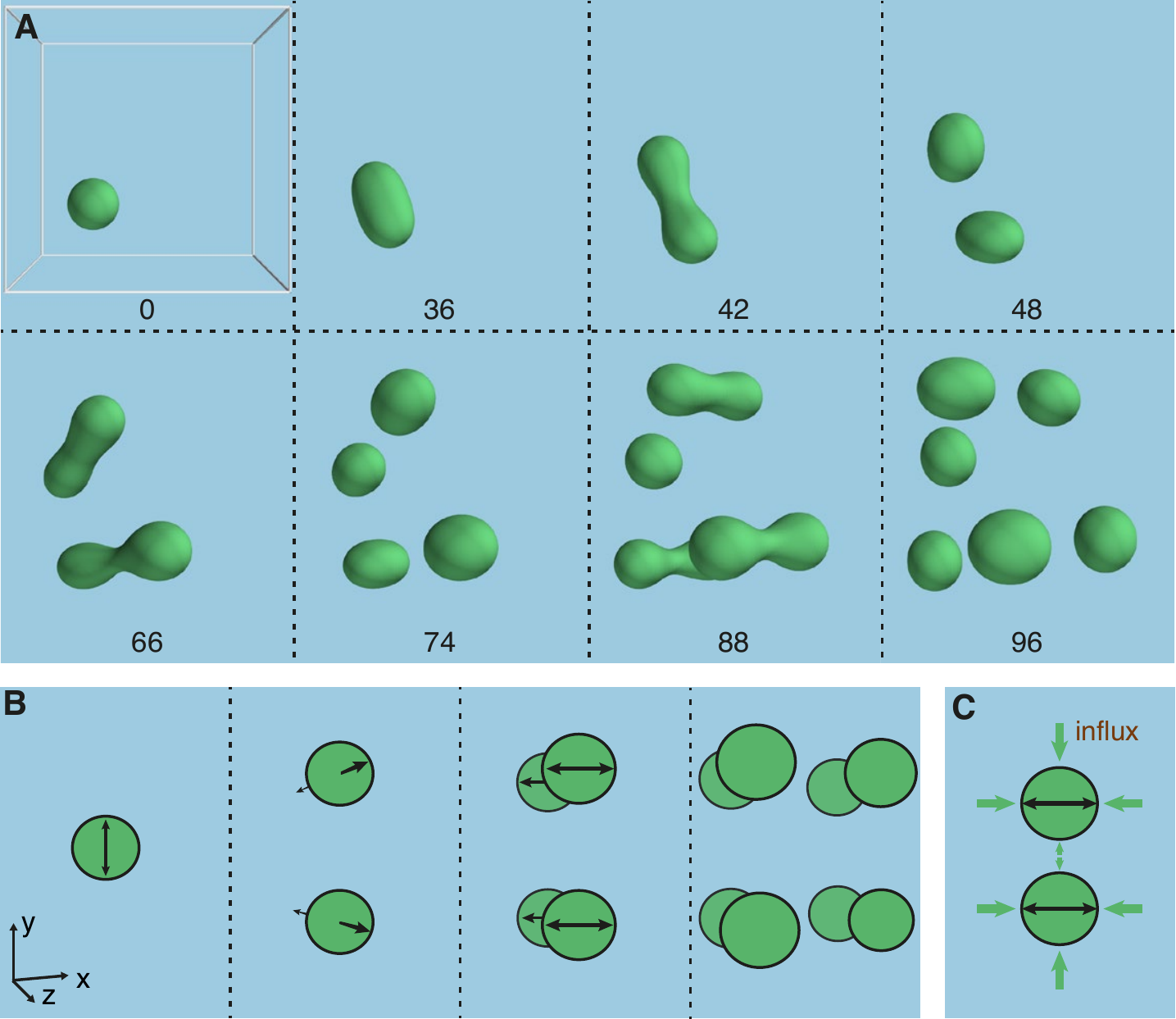}
	\caption[Multiple divisions]{
	A) Sequence of droplet divisions at different times as indicated.
	Droplet configurations obtained from numerical solutions to
	the continuum model are represented as three dimensional
	shapes. Parameter $\nu_+ t_0 / \Delta c= 2 \cdot 10^{-3}$. 
	Remaining parameters are the same as in Fig. 1C. 
	B) Schematic representation of the orientation of subsequent
	division axes. 
	C) Droplet division is oriented along the axis
	for which diffusion fluxes (orange arrows) are maximal.
	}
	\label{fig:cycle}
\end{figure}

In order to study how the shape instability leads to droplet division, 
we investigated the droplet dynamics beyond the linearized analysis
using the continuum model. This model can capture the topological
changes of the droplet surface that occur during division.
Numerical calculations of the continuum model
 (see supplemental information) confirm 
the results of the stability analysis. 
An example of droplet division is shown in \figref{fig:snapshots}C.
The state diagram for the continuum model 
is shown in \figref{fig:stability_diagram}C. Comparing the state diagrams
\figref{fig:stability_diagram}B and \figref{fig:stability_diagram}C reveals that both models exhibit qualitatively the same behaviors. Note that due to simplifications 
 in the effective droplet model, the parameters are different in both models  
(see supplemental information)  and the regions in both diagrams differ slightly. 
 While \figref{fig:stability_diagram}B only shows where
droplets become unstable (red line), \figref{fig:stability_diagram}C reveals the behaviors 
of droplets in the unstable region.  We find that droplets typically divide into two daughters
(red circles). However, for some parameter values they divide into three droplets (red triangles).
 Our calculations show that droplets typically divide after they become unstable.
 However, in some cases division was not seen during the time of calculations (red rectangles). 
In these cases droplets elongated until they reaches the size of the simulation box. It is unclear
whether they would divide in a larger box. 

Our numerical calculations also reveal that droplets 
typically undergo multiple divisions, see \figref{fig:cycle}A and supplemental movie. 
After a first division, the smaller daughters
grow until they divide again when they reach the radius $R_{\rm div}$. 
Interestingly, the division axes are  not independent of each other, see \figref{fig:cycle}A. 
In the absence of system boundaries,
the division axes of both daughters 
are perpendicular to the first division axis, see \figref{fig:cycle}B.
Similarly, when the four granddaughters divide, their division axes are perpendicular to
both the division axes of the first and the second division. 
The division axes in subsequent droplet divisions are determined by
droplet interactions via the concentration fields surrounding the droplets.
The two growing
daughter droplets effectively compete for droplet material, leading to the
depletion of droplet material in the space between them. 
Therefore, diffusion fluxes and 
growth rates are larger along axes 
perpendicular to the previous division axis, see \figref{fig:cycle}C. This bias due to droplet
interactions determines the division axes. In our numerical calculations, boundary conditions
also influence the droplet divisions and slightly modify the division axes, see \figref{fig:cycle}A.

\paragraph*{Discussion}

The question how life first arose on earth has fascinated both scientists and nonscientists
since it was understood that modern life emerged by evolution from early precursors.
While evolution can  be reconstructed to a large extend both from fossil records
and from the phylogenetic analysis of todays genomes, the structure and nature of 
early life forms remain quite unclear~\cite{Woese1990}. 
How did the first replicating cells
emerge from prebiotic precursors?  Since replication involves specific chemical reactions,  
early replicators had to spatially organize
chemistry and to concentrate certain molecules to facilitate reactions
that would be unlikely in dilute or disorganized situations. Therefore, protocells 
as containers for chemical reactions had to appear.

Alexander Oparin pioneered the idea that macromolecular aggregation could lead to the
formation of 'coacervates', liquid 
droplets that could organize chemistry and provide microreactors in which selected
molecules were concentrated for
prebiotic chemistry~\cite{Oparin1952,Fox1976}. What types of molecules could have formed such droplets?
It is interesting to note that modern day cells possess a number of chemical compartments 
that are not separated by a membrane from the cell 
cytoplasm but that form by phase separation from the cytoplasm~%
\cite{Brangwynne2009,Brangwynne2011a,Toretsky2014,Hyman2014}.
Many of these compartments are liquid and consist of RNA molecules 
and RNA binding proteins~\cite{Weber2012,Elbaum-Garfinkle2015,Molliex2015,Lin2015}. 
The RNA world hypothesis suggests that at the origin of life,
RNA was both the carrier of genetic information and could have acted as early enzymes~\cite{Gilbert1986,Higgs2015}. 
Folded RNA molecules called ribozymes can be catalysts for many reactions including RNA processing~\cite{Fedor2005}. 
Combining RNA with other molecules such as simple peptides may have been 
sufficient to organize RNA in liquid droplets \cite{Koga2011}.

The steps from chemically active droplets to the first dividing cells with membranes  
pose a big challenge to the understanding of early evolution.  
While it has been suggested that ribozymes that replicate
RNA could have formed by molecular evolution~\cite{Unrau1998,Higgs2015}, it is unclear how a cell membrane and cell division could have emerged \cite{Hanczyc2003,Hanczyc2004,Macia2007,Murtas2013}. 

The possibility that droplets may spontaneously divide has been discussed in the
context of either negative surface tension \cite{Browne2010,Patashinski2012} or in active nematic droplets \cite{Giomi2014}.
Here we show that simply adding a proto-metabolism to droplets formed by classical phase separation
can naturally lead to droplet division despite their surface tension. 
Membranes or surfactants are therefore not required to achieve 
division of prebiotic cells.
Active droplets are natural systems to organize 
the chemistry of replicators and to
form protocells. 
Such droplets can in principle 
form spontaneously by a rare nucleation event. Once they exist, they grow and divide. 
They provide a container for chemical reactions and they concentrate selected molecules 
that have an affinity to the droplet phase. 
The liquid and dynamic nature of active droplets 
implies that components in the droplet can mix and chemical reactions are facilitated.
Protocells formed by active droplets require a constant energy supply, which could have
been provided by a chemical fuel, by tides, or by temperature gradients, 
e.g. in hydrothermal vents 
on the sea floor~\cite{haldane1929origin,baross1985,Martin2012,Martin2014}. The chemical reactions by which new droplet material is
formed and subsequently degraded represents an early metabolism.

The fact that active droplets tend to become unstable and divide is a very unusual
behavior of droplets. Usually,  
droplets maintain their spherical shape because surface
tension tends to reduce the surface area. An instability of the droplet shape requires 
non-equilibrium conditions. 
In our models, the chemically driven diffusion fluxes 
associated with stationary droplets trigger the shape 
instability. In the absence of chemical reactions and stationary fluxes,
the shape instability does not occur. 
This is reminiscent of other well known shape instabilities of moving interfaces.
The droplet instability discussed here that triggers droplet division is related
to the Mullins-Sekerka instability often discussed in the context of crystal growth~\cite{Mullins1963}. In the
case of the Mullins-Sekerka instability, an interface advances because of a diffusive influx. Beyond a critical interface velocity, a flat
interface becomes unstable with respect to growing spikes called dendrites.
The Mullins-Sekerka instability occurs for a moving interface 
in the absence of chemical reactions, while the active droplet instability
discussed here requires reactions but no interface motion. 
In the limit of large characteristic
length scales introduced by the chemical reactions, the
conditions for both instabilities become the same (supplemental information).

We propose that active droplets that are maintained away from thermodynamic
equilibrium by a constant influx of nutrients and a constant efflux of waste are 
a simple model of membraneless protocells. One can speculate that such droplets
could have concentrated RNA molecules together with other molecular species
to form early replicators with an early metabolism. It is interesting to envision early ecosystems in which
droplets of different type may have had symbiotic relationships if one produces the
nutrient of the other. Alternatively one can find predator-prey relationships when a 
droplet fuses with a different one to harvests its resources. 

Finally, the possibility that early protocells
were active droplets suggest possible scenarios by which cell membranes and cells with
a more modern architecture could have
emerged. The droplet surface is an interface that will in general attract amphiphilic molecules.
Such molecules 
have neither an affinity for the droplet phase nor for the surrounding fluid. As a result,
selected molecules populate the droplet surface and surface chemical reactions could
be established. Such surface modifications could improve the resistance of droplets to
varying 
environmental conditions and provide specific surface properties. If lipids were available in the 
outside fluid,
lipid bilayers could be attracted to the specific droplet surface chemistry.  Our work shows
that active droplets can naturally divide. Therefore protocells could have obtained their membranes
long after the first dividing cells had appeared on earth.

\nocite{atkins2010atkins}
\nocite{Desai2009}
\nocite{Cahn1958}
\nocite{Glotzer1994,Christensen1996}
\nocite{Cahn1958}
\nocite{Dennis2013}
\nocite{Bray1994}
\nocite{Zwicker2015}
\nocite{Zhong-can1987}
\nocite{Mullins1963}

\bibliographystyle{my_naturemag_noURL}
\bibliography{papers,ool,ool2,RNA_drops,RNA_drops2}
\clearpage



**Box 1: Reaction rates and energy supply**

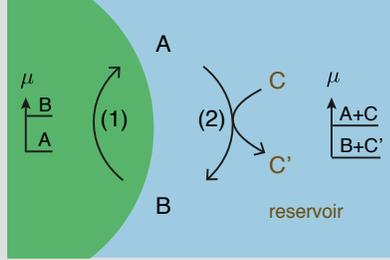

Figure 4: Schematic representation of the reaction cycle involving two pathways (1) and (2). The differences of the chemical potentials $\mu$ determine the direction of the spontaneous reactions: Coupling to the chemical fuel $C$ with reaction product $C'$ drives pathway (2) in the direction $A \to B$ outside the droplet, while $B \to A$ is occurring in the absence of $C$ inside the droplet.

The chemical reaction $A \rightleftharpoons B$ converts soluble precursors $A$ to droplet material $B$ with forward reaction flux $s_\to$ and reverse flux $s_\leftarrow$. The net reaction flux $s = s_\to - s_\leftarrow$ characterizes the concentration per unit time that is undergoing the reaction. Compatibility with thermodynamics requires (45)

$$\frac{s_\to}{s_\leftarrow} = \exp\left(-\frac{\Delta\mu}{k_\mathrm{B}T}\right) \quad , \tag{B.1}$$

where $\Delta\mu$ is the chemical free energy change associated with the forward reaction. This condition leads to detailed balance of forward and backward reaction rates at chemical equilibrium. The net reaction flux $s$ can therefore be written as

$$s = s_\leftarrow \cdot \left[\exp\left(-\frac{\Delta\mu}{k_\mathrm{B}T}\right) - 1\right] \quad . \tag{B.2}$$

Chemical equilibrium is reached when $\Delta\mu = 0$ and the net reaction flux vanishes, $s = 0$. If as in (1) the reaction does not involve other reaction partners or external energy input, the chemical free energy change $\Delta\mu = \Delta\mu^{(1)}$ is given by the difference of the chemical potentials,

$$\Delta\mu^{(1)} = \mu_B - \mu_A \quad . \tag{B.3}$$

Such a reaction leads to spontaneous degradation of $B$ and formation of $A$ if $\mu_B > \mu_A$ and thus $\Delta\mu^{(1)} > 0$. The chemical potentials of a molecular species $n$ can be written as $\mu_n = k_\mathrm{B}T\ln(v_n c_n) + w_n$, where $v_n$ is a molecular volume and $c_n$ the concentration of species $n$. The first term is of entropic origin while the contribution $w_n$ is mainly enthalpic and includes internal molecular free energies and interaction energies between molecules. Note that $w_n$ generally depends on composition.

The net reaction rate corresponding to reaction pathway (1) can thus be written as

$$s^{(1)} = s_\leftarrow^{(1)} \cdot \left(\frac{c_A}{c_B}K^{(1)} - 1\right) \tag{B.4}$$

where $K^{(1)} = (v_A/v_B)\exp((w_A - w_B)/k_\mathrm{B}T)$ is the equilibrium constant of reaction pathway (1). Note that in the case of phase separation, this equilibrium constant can have different values inside and outside the droplet. If only reaction pathway (1) occurs, droplets are passive despite the presence of the reaction and the system reaches a thermodynamic equilibrium. No droplet divisions occur. Such a system exhibits Ostwald ripening and after long times reaches an equilibrium which contains either a single large droplet or no droplet.

Active droplets require an external energy supply that maintains the droplets away from thermodynamic equilibrium at all times. The reaction $A \rightleftharpoons B$ can be coupled to an externally supplied fuel $C$ with reaction product $C'$ with chemical potential difference $\Delta\mu_C = \mu_C - \mu_{C'} > 0$. This second reaction pathway (2) obeys Eq. (B.2) with $\Delta\mu = \Delta\mu^{(2)}$ and

$$\Delta\mu^{(2)} = \mu_B - \mu_A - \Delta\mu_C \quad . \tag{B.5}$$

The corresponding reaction flux can be written as

$$s^{(2)} = s_\leftarrow^{(2)} \left(\frac{c_A}{c_B}K^{(2)} - 1\right) \tag{B.6}$$

with equilibrium constant $K^{(2)} = K^{(1)}\exp(\Delta\mu_C/k_\mathrm{B}T)$. If both pathways are active at the same time, the net reaction flux is $s = s^{(1)} + s^{(2)}$. In this paper we consider the case where an active droplet converts $B$ to $A$ inside the droplet mainly via the reaction pathway (1) while outside the droplet material $A$ is used to generate $B$ mainly via the reaction pathway (2) using the external fuel as an energy source, see Fig. 4. No chemical equilibrium can be reached in this case because the equilibrium constants $K^{(1)}$ and $K^{(2)}$ imply incompatible equilibrium conditions. The droplet is thus active.

**Box 2: Dynamics of active droplets**

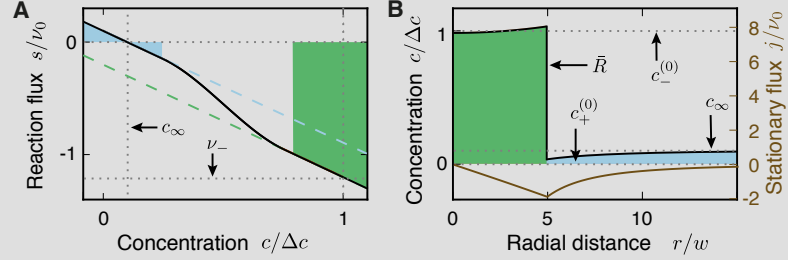

Figure 5: A) Chemical reaction flux as a function of concentration (solid). The linearized flux inside and outside the droplet is indicated as dashed lines. B) Stationary concentration profile of an active droplet (orange) with radius $\bar{R}$ in background fluid (blue). The equilibrium concentrations $c_\pm^{(0)}$ and the concentration far from the droplet $c_\infty$ are indicated. The green line (axis on the right) shows the stationary flux $j = -D_\pm \partial_r c$. Parameter values are the same as in Fig. 2A, with $\nu_-/\nu_0 = 1.2$, $\nu_+/\nu_0 = 0.1$.

We consider a fluid that contains the droplet forming material $B$ at concentration $c = c_B$. The system segregates into two coexisting phases that are separated by a sharp interface. We consider the limit of strong segregation of phases by a sharp interface. Across the interface, chemical potentials are continuous, $\mu_+ = \mu_-$, while the pressure exhibits a jump

$$P_- - P_+ = 2\gamma H \quad, \tag{B.7}$$

known as Laplace pressure. Here, $\gamma$ denotes surface tension and $H$ the local mean curvature of the interface. The subscripts $-$ and $+$ refer to values at the interface inside and outside the droplet, respectively. These thermodynamic conditions determine the concentrations $c_-$ and $c_+$ at the interface where both phases coexist. Because the Laplace pressure depends on local curvature, the equilibrium concentrations also depend on curvature $H$. We express this dependence to linear order by

$$c_\pm \simeq c_\pm^{(0)} + \gamma \beta_\pm H \,, \tag{B.8}$$

where $c_\pm^{(0)}$ denote the equilibrium concentrations of coexisting phases at a flat interface and we have introduced the coefficients $\beta_\pm$ to describe the effects of interface curvature.

The droplet material $B$ is produced by chemical reactions with total reaction flux $s$, which is a function of concentration, see Fig. 5A and Box 1. The time-evolution of the concentration field $c$ is then described by the reaction-diffusion equation

$$\partial_t c = D_\pm \nabla^2 c + s \,, \tag{B.9}$$

where $D_+$ and $D_-$ denote the diffusion coefficients outside and inside the droplet, respectively. The evolution of the droplet shape is governed by the normal velocity of the droplet interface

$$v_n = \frac{j_- - j_+}{c_- - c_+} \,, \tag{B.10}$$

where $j_\pm = -\mathbf{n} \cdot D_\pm \nabla c$ are the normal diffusion fluxes at the interface, inside and outside the droplet, and $\mathbf{n}$ denotes the surface normal.

The reaction flux is typically positive ($B$ is produced) outside the droplet, while it is negative ($A$ is produced) inside, see Fig. 5A. We expand the function $s(c)$ in terms of the concentration variations inside and outside the droplet to linear order as

$$s(c) \simeq \begin{cases} \nu_+ - k_+\bigl(c - c_+^{(0)}\bigr) & \text{outside the droplet} \\ -\nu_- - k_-\bigl(c - c_-^{(0)}\bigr) & \text{inside} \end{cases}. \tag{B.11}$$

The reaction rates $k_\pm$ characterize the relaxation times inside and outside the droplet. The flux of production of $B$ molecules at the equilibrium concentrations outside and inside the droplet are denoted $\nu_+$ and $\nu_-$, respectively. We call $\nu_-$ turnover because it is the flux at which $B$ molecules disappear inside the droplet. The concentration field varies over the characteristic length scales $\ell_\pm = (D_\pm/k_\pm)^{1/2}$ inside and outside the droplet, respectively.

At large distances $r \gg \ell_+$ from the droplet, the net reaction flux $s(c)$ vanishes and the concentration reaches the constant value $c_\infty = c_+^{(0)} + \nu_+/k_+$. The chemical reactions thus generate a supersaturation

$$\epsilon = \frac{c_\infty - c_+^{(0)}}{\Delta c} \,, \tag{B.12}$$

where $\Delta c = c_-^{(0)} - c_+^{(0)}$. This supersaturation drives the diffusion flux $j_+$ toward the droplet interface. Inside the droplet, droplet material is degraded, leading to a concentration profile with minimal concentration in the droplet center. This causes a diffusion flux $j_-$ towards the center, see Fig. 5B.


\setcounter{figure}{0} 
\setcounter{equation}{0} 
\renewcommand{\theequation}{S.\arabic{equation}}
\renewcommand{\thefigure}{S\arabic{figure}}


\renewcommand{\thesection}{\Roman{section}}

{\large \begin{center}
Supplement \\
\centering Growth and Division of Active Droplets: A Model for Protocells
\end{center}}


\section{Continuum model for active droplets}

\subsection{Free energy function and chemical rates}
\label{sec:continuous_model}

We consider an incompressible fluid 
containing two components: a component $A$ that forms 
the background fluid and a droplet material $B$ that forms droplets by phase separation.
Chemical reactions  convert the two components 
into each other. 

The concentration of the droplet material $B$ 
is denoted by $c(\vect r, t)$ where ~$\vect r$
is the position and $t$ denotes time.
The concentration of the second component can be determined 
from $c$ using the incompressibility condition. Therefore,  
the free energy density~$f$ only depends on the concentration $c$.
We use the following double-well free energy function
\begin{equation}
	f(c) =
	 \frac{b}{2 (\Delta c)^2}
			\Bigl(c- c^{(0)}_-\Bigr)^2
			\Bigl(c- c^{(0)}_+\Bigr)^2
	\label{eqn:free_energy_density}
	\;,
\end{equation}
where we have defined $\Delta c = \bigl| c^{(0)}_- - c^{(0)}_+ \bigr|$.
Here, the positive parameter $b$ characterizes  
molecular interactions and entropic contributions.
This free energy describes 
the segregation of the fluid 
in two coexisting 
phases~\cite{Desai2009}: one phase rich in droplet material with $c \approx c^{(0)}_-$ and a diluted phase with $c \approx c^{(0)}_+$.

The state of the system is characterized by the free energy 
\begin{equation}
	F[c] = \int \Bigl[
		f(c)
		+ \frac{\kappa}{2} \bigl(\Nabla c \bigr)^2
	\Bigr] \diff^3 r
	\label{eqn:free_energy}
	\;,
\end{equation}
where the integral is over the system volume.
Here, the coefficient $\kappa$ 
is related to surface tension and the interface width~\cite{Cahn1958}.
The chemical potential $\bar\mu = \delta F[c]/\delta c$, which governs demixing,
reads
\begin{equation}
	\bar\mu =  
		\frac{b}{(\Delta c)^2}
			\bigl(c- c^{(0)}_+\bigr)
			\bigl(c- c^{(0)}_-\bigr)
			\bigl(2c- c^{(0)}_- - c^{(0)}_+\bigr)
		- \kappa\Nabla^2 c
	\; .
	\label{eqn:chemical_potential}
\end{equation}
The dynamics of the concentration field is described by the reaction-diffusion equation~\cite{Glotzer1994, Christensen1996}
\begin{equation}
	\partial_t c = m \Nabla^2 \bar \mu + s(c)
	\label{eqn:cahn_hilliard}
	\; .
\end{equation}
Here, $m$ is a mobility coefficient of the droplet material. 
The source term $s(c)$ describes chemical reactions.

We choose the function $s(c)$ to be linear in the phases
outside and inside the droplet. We connect these linear behaviors 
by a cubic interpolating polynomial:
\begin{equation}
 s(c) = 
 \begin{cases}
 \nu_+ + k_+ c_+^{(0)} + k_+ c  & \text{for } c < c_{\rm c}^+ \\
 \nu_- + k_- c_-^{(0)}  + k_- c  & \text{for } c > c_{\rm c}^- \\
 p(c) & \text{for }   c_{\rm c}^+<c<c_{\rm c}^-
 \end{cases} \; ,
 	\label{eqn:source}
\end{equation}
where $c_{\rm c}^+$ and $c_{\rm c}^-$ are two characteristic concentrations
and $p(c) = a_0 + a_1 c + a_2 c^2 + a_3 c^3$ is a cubic polynomial. 
The coefficients $a_i$ are
determined uniquely
by the conditions that $s(c)$ and its derivative are continuous functions: 
\begin{linenomath*}
\begin{subequations}
\label{eqn:conditions}
\begin{align}
p(c_{\rm c}^+) &= \nu_+ + k_+ c_+^{(0)} + k_+ c_{\rm c}^+  \\[1pt]
p(c_{\rm c}^-) &= \nu_- + k_- c_-^{(0)} + k_- c_{\rm c}^-  \\[1pt]
p'(c_{\rm c}^+) &= k_+  \\[1pt]
p'(c_{\rm c}^-) &= k_-  \quad .
\end{align}
\end{subequations}
\end{linenomath*}
The reaction flux given in \Eqref{eqn:source}
describes a situation where an external energy source
drives the system away from equilibrium, see Box 1.  \Eqsref{eqn:chemical_potential}--\eqref{eqn:source} 
define the continuum model of active droplets.

\subsection{Relation with the effective droplet model}
\label{sec:effective_model}
The model described by \Eqref{eqn:cahn_hilliard} typically forms distinct phases, 
which are separated by an interface.
Considering a flat interface between two phases with bulk concentrations 
\mbox{$c= c^{(0)}_-$} and $c= c^{(0)}_+$, 
the free energy $F$ given in \Eqref{eqn:free_energy} is minimized by
the concentration profile
\begin{equation}
	c^*(x) = 
	\frac{c^{(0)}_- + c^{(0)}_+}{2} + \frac{c^{(0)}_- - c^{(0)}_+}{2}\tanh\frac{x}{w}
	\;,
\end{equation}
where $x$ is a coordinate that is normal to the interface and $w=2(\kappa/b)^{1/2}$ is the interface width~\cite{Cahn1958}.
The surface tension, i.e. the free energy per unit area of the interface, is~\cite{Desai2009}
\begin{equation}
	\gamma = \int_{-\infty}^{\infty} F[c^*(x)] \diff x
		= \frac{(\Delta c)^2}{6} \sqrt{\kappa b}
	\;.
\end{equation}
Two different bulk concentrations~$c_-$ and $c_+$ coexist across the interface for
which the chemical potential is equal on both sides. For a curved interface the
pressure difference between the inside and outside of the droplet is  
the Laplace pressure $2\gamma H$, where $H$ is the mean curvature of the interface.
These two equilibrium conditions read
\begin{linenomath*}
\begin{subequations}
\label{eqn:ternary_fluid_equilibrium}
\begin{align}
	0 &= \bar \mu(c_-) - \bar \mu(c_+)
	\label{eqn:ternary_fluid_equilibrium_mu}
\\
	0 &= (c_- - c_+ )\bar \mu(c_-)
	 + f(c_+) - f(c_-)  - 2\gamma H
	\label{eqn:ternary_fluid_equilibrium_pressure} \;,
\end{align}
\end{subequations}
\end{linenomath*}
where $c_-$ and $c_+$ denote the concentration at the interface inside and outside the droplet, respectively.
Using the free energy density as defined in \Eqref{eqn:free_energy_density}, the concentrations that obey \Eqsref{eqn:ternary_fluid_equilibrium} can be expressed 
to first order in~$H$ as
\begin{linenomath*}
\begin{subequations}
\label{eqn:phase_separation_result}
\begin{align}
	c_- &\approx c^{(0)}_- + \beta \gamma H \\
	c_+ &\approx c^{(0)}_+ + \beta \gamma H
	 \;,
\end{align}
\end{subequations}
\end{linenomath*}
which is valid for small surface tension, $\gamma \ll R\Delta c / \beta$.
Here, the coefficient $\beta = 2/(b\Delta c)$ describes the effect of Laplace pressure on the concentration at the interface.
Note that $\gamma\beta$ defines a length scale, which is related to the interface width by~$\gamma\beta =  w\Delta c/6$.
Linearizing \Eqref{eqn:cahn_hilliard} at the values $c_+^{(0)}$ and $c_-^{(0)}$ outside and inside the droplet 
gives the linear reaction-diffusion equation defined in Box~2, with diffusivity $D = m b$.

We thus can relate 
the parameters $b$, $\kappa$, and $m$ of the continuous theory to the 
parameters $\gamma$, $\beta_\pm$, and $D_\pm$ of the effective droplet model.
In particular, $\beta_+ = \beta_- = \beta$, and $D_+ = D_- = D$.

\subsection{Numerical methods}
We solved \Eqref{eqn:cahn_hilliard} with \eqref{eqn:source} and \eqref{eqn:chemical_potential} 
numerically using the xmds2 software package (version 2.2.2)~\cite{Dennis2013}
using an adaptive Runge-Kutta scheme of order 4/5, with tolerance $10^{-5}$.
The Laplace operator was evaluated by a spectral method, 
while the chemical rates were evaluated directly.
Numerical calculations were performed in
a finite volume with no flux boundary conditions. 

We normalize concentration, length and time by 
$\Delta c = c_-^{(0)} - c_+^{(0)}$, $w$ and $t_0 = w^2/D$, respectively, 
where the characteristic length scale is
$w = 2(\kappa/b)^{1/2}$.
The relevant dimensionless model 
parameters are 
$c_\pm^{(0)} / \Delta c$, $k_\pm t_0 $, 
$\nu_\pm t_0/  \Delta c$ and $c_c^\pm / \Delta c$. 
In all numerical calculations, we chose 
$c_+^{(0)} / \Delta c = 0 $, $c_-^{(0)} / \Delta c   = 1$ and $k_\pm t_0 = 10^{-2}$. 


\subsubsection{Stability diagram}
Using three dimensional calculations in Cartesian coordinates, we 
observed that droplet configurations during the division of isolated single droplets were approximately 
axisymmetric.
To determine the stability diagram shown in Fig. 2C we therefore performed calculations in cylindrical coordinates
imposing axisymmetry.
We used an axisymmetric cylindrical box with length 
$60w$ and radius $30w$, 
discretized with $120$ and $60$ points, respectively. 

The initial conditions were given by a concentration profile that corresponded to
a droplet geometry of a slightly prolate ellipsoid 
with unequal half axes with length $R/w - 0.1$ and $R/w + 0.1$,
centered at the box center. The initial droplet size was chosen close to the 
stationary size in the continuum model. As an estimate for the stationary size we typically chose
$R/w = 0.9 \bar R_s / \hat w$.
Here, $\bar R_s$ is the stationary radius calculated in the effective droplet 
model and $\hat w=6 \beta_+ \gamma / \Delta c$, see Section \ref{sec:effective_model}. 
The concentration field at positions $\vect{r}$ was initialized by the function
\begin{equation}
c(\vect r) = \frac{c_\infty + c_-^{(0)}}{2} + \frac{c_\infty - c_-^{(0)}}{2} \tanh \frac{d(\vect r)}{w} \; . \label{initial} 
\end{equation}
where $d(\vect r)$ is the oriented distance of $\vect r$ 
to the nearest point on the ellipsoid. The value of
$d(\vect r)$ is negative for points inside the droplet and positive for points outside.
The concentration far from the droplet is $c_\infty = \nu_{+} / k_+ + c^{(0)}_+$.

We calculated the dynamics of the concentration field
over a time interval $T / t_0 =10^4$, for different values of $\nu_\pm t_0/  \Delta c$.
The parameters $c_c^\pm$ related to the chemical reaction in \Eqref{eqn:source} 
were chosen as $c_{\rm c}^+ / \Delta c = 0.25$ and $c_{\rm c}^- / \Delta c = 0.75$.
Because close to the shape instability the dynamics slows down, 
we may slightly overestimate the region of stability, since we cannot detect the exact instability with the finite time intervals simulated. Contours shown in Fig.~1C
correspond to $c / \Delta c = 0.5$.

\subsubsection{Calculations for multiple divisions}

Several subsequent divisions break cylindrical symmetry. 
The calculations shown in
Fig. 3A were therefore performed in three dimensions using cartesian coordinates. 
We chose a cubic box with side length $L = 50w$ and an equidistant 
discretization of 100 points along each dimension.

Initial conditions corresponded to a spherical 
droplet centered at  $\vect{r}=(L/4, L/4, L/4)$.
The concentration field was initialized with 
$c = c_-^{(0)}$ inside the droplet and 
$c = c_\infty $ outside. 
The parameters for the calculations were $\nu_-  t_0/  \Delta c = 7 \cdot 10^{-3}$, 
$\nu_+ t_0 / \Delta c = 2 \cdot 10^{-3}$ and $c_{\rm c}^+/\Delta c = 
c_{\rm c}^-/\Delta c = 0.5$. Surfaces shown in Fig. 3A
correspond to $c / \Delta c = 0.5$.

\section{Effective model for active droplets}

Using the effective droplet model defined in Box 2,  we discuss steady state droplets 
and perform a linear stability analysis of the spherical droplet shape.  
We determine conditions for a shape instability towards an elongated shape.   

\subsection{Droplet dynamics in spherical coordinates}
Using spherical coordinates $r,\theta,\phi$ centered on the droplet, the interface defining the  droplet surface is positioned at radial distance $r=R(\theta,\phi)$. 
Away from the interface, the concentration field $c(r,\theta,\phi)$ obeys the reaction-diffusion equation 
\begin{equation}
\partial_t  c  =  D_\pm \nabla^{2}  c +s \; . \label{eq:linearRDE}
\end{equation}
Here, $t$ denotes time and $D_+$, $D_-$ are the 
diffusion coefficients outside ($r>R(\theta, \phi)$) and inside ($r<R(\theta, \phi)$) the droplet,
respectively. 
The reaction flux $s$ is given by
\begin{equation}
s = 
\begin{cases}
\nu_+  - k_+(c_+- c^{(0)}_+)  & {\rm for} \quad r>R\\
-\nu_-  - k_-(c- c^{(0)}_-)   & {\rm for }\quad  r<R\\
\end{cases}
\; .
\end{equation}
Here, reaction rates inside and outside the droplet are denoted by $k_\pm$, 
$c^{(0)}_\pm$ denote the equilibrium bulk 
concentrations of coexisting phases near a planar interface. The reaction fluxes 
at equilibrium concentrations are denoted by $\nu_\pm$.
For the concentration $c=c_0$, with 
$c_0 = -\nu_{-} / k_- + c^{(0)}_-$, the reaction flux  
inside the droplet vanishes, while for $c=c_\infty$ with 
$c_\infty = \nu_{+} / k_+ + c^{(0)}_+$ the reaction flux outside the droplet vanishes.

At the interface at $r=R(\theta, \phi)$ we  impose 
boundary conditions for the concentration:
\begin{equation}
c (R_\pm) = c^{(0)}_\pm +  \beta_\pm \gamma H(\theta, \phi) \; . \label{eq:bc}
\end{equation}
This boundary condition describes a concentration jump at the interface.
It  corresponds to local thermodynamic equilibrium at a curved interface with surface tension $\gamma$. Here, $R_\pm $ denote the limits of approaching the interface at radial distance $R(\theta,\phi)$ 
from the outside or the inside, respectively.
The mean curvature of the interface is denoted $H$ and the coefficients
$\beta_\pm$ describe the change of the equilibrium concentration at the interface 
due to Laplace pressure $2\gamma H$.

The normal velocity~$v_n$ of the interface is proportional to the difference of normal fluxes inside and outside \cite{Bray1994}, 
\begin{equation}
	v_n = \vect{n} \cdot \frac{\vect{j}_- - \vect{j}_+}{c(R_-) - c(R_+)}
	\;,
\label{eq:vn}
\end{equation}
with flux $\vect{j}_\pm = -D_\pm \vect \nabla c(R_\pm)$ and unit vector $\vect n$ normal to the interface.
The droplet shape $\vect{R}(\theta,\phi)=R(\theta,\phi) \vect{e}_r$, 
where $\vect{e}_r$ denotes the unit vector in radial direction,
can be parameterized using the angles $\theta$ and $\phi$.
The interface velocity can be written as
\begin{equation}
	\frac{\partial \vect R(\theta,\phi,t)}{\partial t}
	= v_\theta \vect e_1 +v_\phi  \vect e_2 + v_n \vect n
	\;,
\end{equation}
where $\vect e_1=\partial \vect R/\partial\theta$ and
 $\vect e_2=\partial \vect R/\partial\phi$ are the two basis vectors of the tangential plane. 
Using $\partial \vect R/\partial t= (\partial R/\partial t )\vect e_r$, 
the velocity components $v_\theta$ and $v_\phi$ can be obtained from the conditions
$(\partial \vect R/\partial t) \vect e_\theta = 0$ and $(\partial \vect R/\partial t) \vect e_\phi = 0$. 
Here, $\vect e_\theta$ and $\vect e_\phi$ are the local normalized basis vectors corresponding to $\theta$ and $\phi$ in spherical coordinates.
The radial interface velocity $\partial R/\partial t=(\partial \vect R/\partial t) \cdot \vect e_r$
then reads
\begin{equation}
\frac{\partial R}{\partial t}  = v_n \left[1 + \left( \frac{\partial_\theta R}{R} \right)^2 + \left( \frac{\partial_\phi R}{R \sin \theta} \right)^2 \right]^{\frac12} \; ,
\label{eq:Rdot}
\end{equation}
where $v_n$ is given by (\ref{eq:vn}).

\subsection{Stationary states of spherical droplets}

\begin{figure}
\centering{}\includegraphics[width=0.9 \columnwidth]{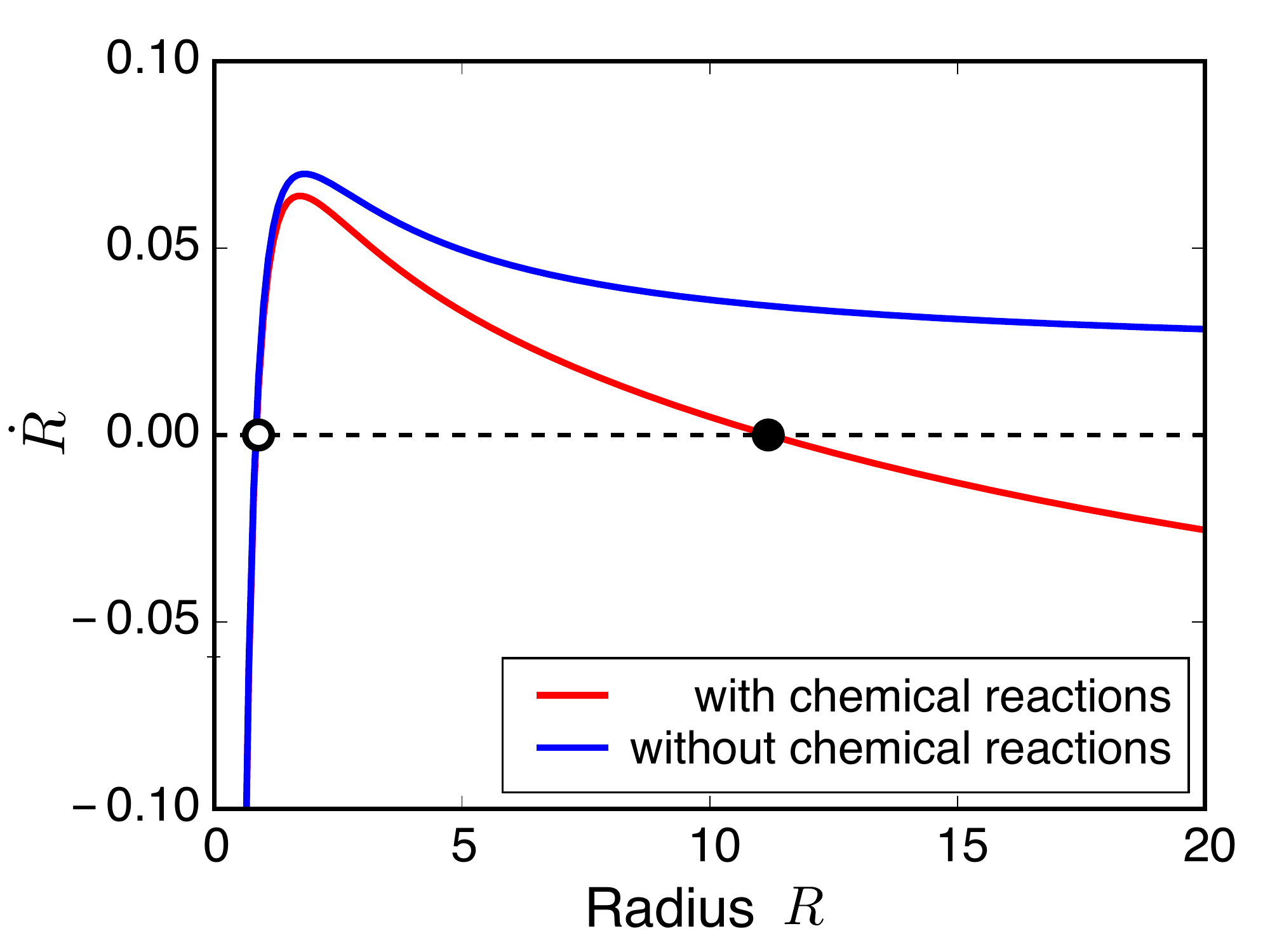}\protect
\caption{
Rate of droplet growth $\diff R/\diff t$ as a function of droplet radius $R$ in a quasistatic limit in the presence of
chemical reactions (red line) and without chemical reactions (blue line).  
The zeros of $\dot{R}$ correspond to stationary radii. An unstable critical radius (white circle)
and a stable droplet radius (black circle) are indicated. 
Parameter values are: $\nu_- \tau_0 / \Delta c =-10^{-2}$ (red line) or $\nu_- \tau_0 / \Delta c =0$ (blue line), $\nu_+ \tau_0 / \Delta c =2 \cdot 10^{-3}$ , 
$k_{\pm} \tau_0 = 0.01$, $c_+^{(0)} =0$, $\beta_- = \beta_+$, $D_- = D_+$.
Here, $\hat w = 6 \beta_+ \gamma / \Delta c$, and $\tau_0 = D_+ /\hat w^2$ are 
characteristic length and time scales. 
\label{fig:Rdot}}
\end{figure}

Stationary solutions to \Eqref{eq:linearRDE} with
 spherically symmetric concentration field
can be expressed as
\begin{equation}
\bar{c} (r) = A_\pm + B_\pm \frac{e^{r / l_\pm}}{r} + C_\pm \frac{e^{ - r / l_\pm }}{r} \; , \label{eq:stat_sol}
\end{equation}
where $l_\pm = (D_\pm / k_\pm )^{1/2}$ are characteristic length scales.
Here, the coefficients $A_\pm$ are set by the chemical reactions,
\begin{equation}
A_\pm = \pm\frac{\nu_{\pm}}{k_\pm } + c^{(0)}_\pm \; . 
\end{equation}
Regular behavior at $r = 0$ implies $C_{-} = -B_{-}$.
For an infinite system, the concentration
far from the droplet reaches a constant value. This implies $B_+=0$.
Using the boundary conditions \eqref{eq:bc} at the interface of a spherical droplet
of radius $R$ we obtain the remaining coefficients
\begin{linenomath*}
\begin{subequations}
\begin{align}
C_{+} &= \left(\frac{\gamma\beta_+ }{R} - \frac{\nu_{+}}{k_+ } \right) R \exp(R / l_+ )
\\[6pt]
B_{-} &= \left(\frac{\gamma\beta_- }{R} + \frac{\nu_{-}}{k_- } \right) \frac{R}{2 \sinh(R/l_- )} \; .
\end{align}
\end{subequations}
\end{linenomath*}
The normal fluxes at the droplet interface are
\begin{linenomath*}
\begin{subequations}
\begin{align}
j_+ (R) & = \frac{D_+ }{R}  \left( \frac{ \gamma \beta_+}{R} - \frac{\nu_{+}}{k_+ }  \right)
	\left( 1 + \frac{R}{l_+ } \right)
\\[6pt]
j_- (R) & = \frac{D_- }{R} \left( \frac{ \gamma \beta_-}{R} + \frac{\nu_{-}}{k_- } \right)
	\left(1 -  \frac{R}{l_- } \coth \frac{R}{l_- }  \right) \; .
\end{align}
\end{subequations}
\end{linenomath*}
Using these steady state fluxes in \Eqref{eq:Rdot} and \Eqref{eq:vn} provides 
a relation between $\diff R/\diff t=v_n$ and the droplet radius $R$ in a quasi-static limit. 
Steady state droplets exist for  
radii $R=\bar R$ for which $\diff R/\diff t$ vanishes. These stationary radii thus obey
\begin{equation}
j_+ (\bar R) = j_- (\bar R)  \; . \label{eq:stability}
\end{equation}
\figref{fig:Rdot} shows an example of $\diff R/\diff t$ as a function of $R$ in the presence (red line) and
absence (blue line) of chemical reactions.
If chemical reactions are present, 
two steady state radii denoted $\bar R_{\rm c}$ (white circle) and $\bar R_{\rm s}$ (black circle)
exist,  corresponding to a critical nucleation radius and a stationary droplet radius, respectively. 
Both stationary radii are shown in Fig. 2A in the main text.

In the limit of large characteristic lengths $l_\pm$ compared to  the droplet radius
$R$,  the stationary radii can be approximated as 
\begin{equation}
\bar R_{\text{c}}\approx\frac{\gamma\beta_+}{c_\infty -c^{(0)}_+ } \; , \label{Rnucl}
\end{equation}
and
\begin{equation}
\bar R_{\text{s}} \approx\sqrt{\frac{3D_+ (c_\infty -c^{(0)}_+ )}{\nu_- }} \; , \label{eq:stat}
\end{equation}
where we have used 
$R\ll l_-$ and $\beta_\pm \gamma / [\bar R(c^{(0)}_- - c^{(0)}_+)] \ll 1$. The latter
is obeyed for a sharp interface, see section \ref{sec:effective_model}.

The critical radius estimated by
\Eqref{Rnucl} is closely related to the classical expression for the 
critical nucleation radius of passive droplets. 
In the case of active droplets, 
the supersaturation $\epsilon = (c_\infty - c^{(0)}_+)/ \Delta c$ 
is determined by chemical reactions instead of the  
amount of material provided.
The stationary droplet radius given in
\Eqref{eq:stat} 
describes an inherently non-equilibrium stationary state that
is maintained by opposing fluxes \cite{Zwicker2015}.

\subsection{Stability analysis of the spherical droplet shape}

To analyze the linear stability of the stationary droplets, 
we linearize the dynamic equations in the vicinity
of the stationary state and identify the dynamic
eigenmodes.  The stationary state is unstable with respect to a
dynamic mode if the corresponding growth rate is positive. 

\subsubsection{Linearization at the stationary solution}

We linearize the dynamic equations \eqref{eq:linearRDE}--\eqref{eq:vn} and \eqref{eq:Rdot}
around a stationary solution $\bar c(r)$, which obeys \Eqsref{eq:stat_sol}--\eqref{eq:stability}.
Introducing small perturbations $\delta c$ and $\delta R$ of the concentration field and the droplet
shape, respectively, we write
\begin{equation}
c(r,\theta,\varphi,t)=\bar{c}(r)+\delta c(r,\theta,\varphi,t) \; ,
\end{equation}
and \begin{equation}
R(\theta,\varphi,t)=\bar{R}+\delta R(\theta,\varphi,t) \; .
\end{equation}
The concentration perturbation then obeys
\begin{equation}
\partial_t \delta c = D_\pm \nabla^2 \delta c - k_\pm \delta c \; . \label{eq:lin1}
\end{equation}
The boundary conditions \eqref{eq:bc} become 
\begin{equation}
\delta c(\bar{R}_\pm) =  \beta_\pm \gamma \delta H - \bar{c}' (\bar{R}_\pm) \delta R \; , \label{eq:linbc1a}
\end{equation}
where $\delta H = H(\bar{R} + \delta R) - H(\bar{R})$. 
Using \Eqsref{eq:vn} and \eqref{eq:Rdot}, the time dependence of the droplet shape perturbation is described to linear
order by
\begin{equation}
(c^{(0)}_- - c^{(0)}_+ ) \partial_{t}\delta R = 
D_+ \partial_{r} \delta c(\bar{R}_+) 
- D_- \partial_{r} \delta c(\bar{R}_-)
+  \left[ D_+ \bar{c}''(\bar{R}_+)
	   - D_- \bar{c}''(\bar{R}_-)
   \right] \delta R \; . 
    \label{eq:linbc2}
\end{equation}

\subsubsection{Dynamic modes and relaxation spectrum}

The linearized dynamics of droplet perturbations near the steady state defines a linear operator
$\mathscr{L}$ by 
\begin{equation}
\partial_t \begin{pmatrix}
\delta c \\ 
\delta R 
\end{pmatrix} = \mathscr{L} 
\begin{pmatrix}
\delta c \\ 
\delta R 
\end{pmatrix}  \; .
\end{equation}
The operator $\mathscr{L}$ has eigenfunctions 
$(c_{i}, R_{i})^\intercal$ with corresponding eigenvalues $\mu_{i}$, where $i$ is the mode index.
These modes obey
\begin{equation}
\mathscr{L} 
\begin{pmatrix}
c_{i} \\ 
R_{i} 
\end{pmatrix} = \mu_{i} 
\begin{pmatrix}
c_{i} \\ 
R_{i} 
\end{pmatrix}\; .
\end{equation}
The linear droplet dynamics can thus be decomposed in
eigenmodes with amplitude~$A_i$ as
\begin{equation}
\begin{pmatrix}
\delta c \\ 
\delta R 
\end{pmatrix}  = \sum_i A_i
\begin{pmatrix}
c_i \\ 
R_i 
\end{pmatrix} e^{\mu_i t }
\;,
\end{equation}
where the sum is over all eigenmodes. Thus, the eigenfunctions of $\mathscr{L}$ correspond to 
dynamic modes of the system. For $\mu_i < 0$, 
the values $-\mu_i $ are relaxation rates. The steady state is stable if all $\mu_i < 0$.

\subsubsection{Determination of eigenmodes}

We  determine the eigenmodes and the spectrum of relaxation rates of
a stationary droplet with radius $\bar R$. Because of the spherically symmetric reference
state, we introduce radial and angular indices $i=(n,m,l)$ and use the ansatz
\begin{equation}
\begin{pmatrix}
c_{nlm}(r,\theta,\phi) \\ 
R_{nlm} (\theta,\phi)
\end{pmatrix}  = 
\begin{pmatrix}
c_{nl}(r) \\ 
\epsilon_{nl}
\end{pmatrix} Y_{lm}(\theta \phi) 
\;,
\end{equation}
where $Y_{lm}$ are spherical harmonics and the corresponding
eigenvalues will be denoted $\mu_{nl}$.
Using \Eqref{eq:lin1} with  $r^2 \nabla^2 Y_{lm}=l(l+1) Y_{lm}$, the radial part of the eigenfunctions obeys
\begin{equation}
\left({1 \over r^2} {\partial \over \partial r}
 r^2 {\partial  \over \partial r}-  (\lambda_{nl}^\pm )^2- \frac{l(l+1)}{r^2} \right) c_{nl}(r) = 0 \; ,
  \label{eq:radialLaplacian}
\end{equation}
where
\begin{equation}
(\lambda^\pm_{nl})^2 = \frac{k_{\pm} + \mu_{nl}}{D_\pm} \; .\label{eq:lambda}
\end{equation} 
The boundary conditions \eqref{eq:linbc1a} at $r=\bar R$ can be written as 
\begin{subequations}
\label{eq:eps_c}
\begin{align}
c_{nl}(\bar{R}_+)& = a_{l}^{+}\epsilon_{nl} \label{eq:eps_cp}\\
c_{nl}(\bar{R}_-) & = a_{l}^{-}\epsilon_{nl} \label{eq:eps_cm}
\end{align}
\end{subequations}
with
\begin{equation}
a_{l}^{\pm} = \gamma \beta_\pm \frac{h_{l}}{\bar{R}^{2}} - \bar{c}' (\bar{R}_\pm ) \; ,
\end{equation}
where \cite{Zhong-can1987} $h_{l}=(l^{2}+l-2)/2$. 
From \Eqsref{eq:eps_c} we obtain
a boundary condition at $r=\bar R$:
\begin{equation}
\frac{c_{nl}(\bar{R}_+)}{c_{nl}(\bar{R}_- )}=\frac{a_{l}^{+}}{a_{l}^{-}} 
\label{cpcm}
\; .
\end{equation}
Using \Eqref{eq:linbc2}, we obtain a second boundary condition
\begin{equation}
\left(c^{(0)}_- -c^{(0)}_+ \right)\mu_{nl}  =
 D_+ \bar{c}''(\bar{R}_+)-D_- \bar{c}''(\bar{R}_-)
 + D_+ a_{l}^{+} \frac{c_{nl}'(\bar{R}_+)}{c_{nl}(\bar{R}_+)}
- D_- a_{l}^{-} \frac{c_{nl}'(\bar{R}_-)}{c_{nl}(\bar{R}_-)}
 \; . 
   \label{eq:stab}  
\end{equation}
The boundary conditions \eqref{cpcm} and \eqref{eq:stab} provide jump conditions for both
the values and the first derivatives of the radial modes $c_{nl}(r)$ at $r=\bar R$. 

\subsubsection{Radial profiles and relaxation rates of dynamic modes}

When solving \Eqref{eq:radialLaplacian} with \Eqref{eq:lambda}
to determine the dynamic modes of the system, 
we have to distinguish the cases
$\mu_{nl}<-k_\pm$ and $\mu_{nl}>-k_\pm$, for which the sign of $(\lambda_{nl}^{\pm})^2$ differs. Near an instability of the droplet shape,
an eigenmode exists for which  $\mu_{nl}$ changes sign. Therefore, 
to discuss this instability, it is  sufficient to consider the case $\mu_{nl} > -k_\pm$.
In this case, $(\lambda_{nl}^{\pm})^2$ is positive and solutions to \Eqref{eq:radialLaplacian} are given by modified spherical 
Bessel functions $k_l(\lambda^\pm_{nl}r)$ and $i_l(\lambda^\pm_{nl}r)$. In order to
obtain solutions that are finite at $r=0$ and which do not diverge for large $r$, we 
have
\begin{linenomath*}
\begin{align} 
c_{nl}(r) & =\begin{cases}
\quad \;\; k_{l}(\lambda_{nl}^{+}r) & {\rm for}\quad  r>\bar{R}\\
C_{nl} \; i_{l}(\lambda_{nl}^{-}r) & {\rm for} \quad r<\bar{R}
\end{cases} \; ,
\end{align}
\end{linenomath*}
where the coefficient $C_{nl}$ is determined by boundary conditions  \eqref{cpcm} as 
\begin{equation}
C_{nl}=\frac{a_l^- k_l(\lambda_{nl}^+ \bar R)}{a_l^+ i_l(\lambda_{nl}^-\bar R)} \;.
\end{equation}
The boundary condition \eqref{eq:stab} becomes
\begin{equation}
\left(c^{(0)}_- -c^{(0)}_+ \right)\mu_{nl} =
D_+ \bar{c}''(\bar{R}_+)-D_- \bar{c}''(\bar{R}_-)
+ D_+ a_{l}^{+} \lambda_{nl}^+ \frac{k_{l}'(\lambda_{nl}^+\bar{R})}{k_{l}(\lambda_{nl}^+\bar{R})}   
- D_- a_{l}^{-} \lambda^-_{nl} \frac{ i_{l}'(\lambda^-_{nl}\bar{R})}{i_{l}(\lambda^-_{nl}\bar{R})}
\; .
   \label{mu} 
\end{equation}
Using $\lambda^\pm_{nl}=((k_{\pm} + \mu_{nl})/D_\pm)^{1/2}$,
\Eqref{mu} becomes an implicit equation for the unknown eigenvalues $\mu_{nl}$.
This equation typically has either no solution or one solution. We identify the largest eigenvalue for given $l$ with $n=1$.

In order to determine the full spectrum $\mu_{nl}$ of eigenmodes, we have to consider the case
$\mu_{nl} < - k_\pm$. We then define $(\lambda^\pm_{nl})^2 = -(k_{\pm} + \mu_{nl})/D_\pm$
and the solutions to \Eqref{eq:radialLaplacian}
are of the form $C^\pm_{nl} j_l(\lambda_{nl}^\pm r)+D^\pm_{nl} y_{l}(\lambda_{nl}^\pm r)$,
where $j_l(z)$ and $y_l(z)$ denote spherical Bessel functions, and
the coefficients $C_{nl}$ and $D_{nl}$ are 
determined by boundary conditions. 
The functions $j_l(z)$
and $y_l(z)$  
behave for large $r$ as $j_l(z)\sim z^{-1}\sin(z-l\pi/2)$ and $y_{l}(z)\sim z^{-1}\cos(z-l\pi/2)$.  
\Eqref{eq:stab} now has an infinite set 
of solutions $\mu_{nl}$ for $n>1$, which we order such that $\mu_{nl}>\mu_{n+1,l}$. 
In an infinite system, the set $\mu_{nl}$ 
approaches a continuous spectrum.

\subsubsection{Instability of stationary spherical droplets}

The droplet shape is unstable if at least one mode with $\mu_{1l}>0$ exists. 
We can obtain a criterion for this instability by using $\mu_{nl}=0$ in \Eqref{mu}. 
This leads to 
\begin{equation}
0 =
D_+ \bar{c}''(\bar{R}_+)-D_- \bar{c}''(\bar{R}_-)
  + \frac{D_+ a_{l}^{+}}{l_+} \frac{k_{l}'(\bar{R}/l_+)}{k_{l}(\bar{R}/l_+)}
- \frac{D_- a_{l}^{-}}{l_-} \frac{i_{l}'(\bar{R}/l_-)}{i_{l}(\bar{R}/l_-)}
	\;,
  \label{mu0} 
\end{equation}
which is a condition for the radius $\bar R$ at which the shape becomes
unstable with respect to a deformation characterized by~$l$.

\begin{figure}
\includegraphics[width = 0.9 \columnwidth]{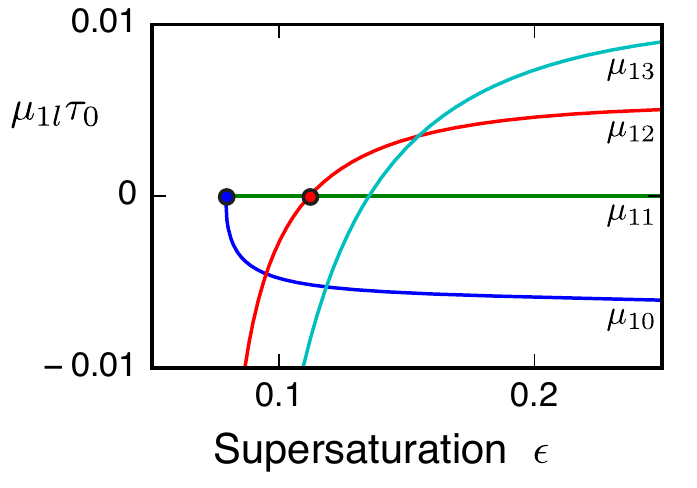}
\caption{
Eigenvalues $\mu_{1l}$ as a function of supersaturation $\epsilon$. At the onset of the instability (red dot) the second mode becomes unstable, leading to droplet division. For larger values of $\epsilon$, higher modes become unstable as well. 
The same parameters as in Fig. 2A (main text), with $\nu_-/\nu_0 = 1$. Stationary and stable radii were used ($\mu_{10} < 0$). 
\label{fig:mu}
}
\end{figure}

Different modes $l$ can become unstable. The case $l=0$ corresponds to changes of the radius.
A droplet with $\mu_{10} < 0$ has a stable radius  $\bar R$. 
For $l=1$ there always exists one marginal mode
with $\mu_{1m}=0$, which corresponds to a translation of the
steady state and does not lead to an instability.  
The first mode that becomes unstable and changes the
droplet shape is the elongation mode $l=2$.

\figref{fig:mu} shows numerically determined values of the largest relaxation rate $\mu_{1l}$
for $l=0,1,2$ and $3$ as a function of supersaturation $\epsilon$ far from the droplet. 
The figure reveals that 
$\mu_{12}$ changes sign and becomes positive as $\epsilon$ is increased, 
indicating the shape instability. 

 \Eqref{mu} can be solved numerically. An approximation of the eigenvalues that is valid in 
 the limit of weak chemical reactions $R \ll l_+$
is
\begin{equation}
\mu_{1l} \simeq 
 (l-1) \frac{D_+}{\Delta c  \bar{R}^{2}}
 \Biggl[
	\bigl(c_\infty -c^{(0)}_+\bigr)
 -\frac{\gamma}{2\bar{R}}
\left(
(4+3l+l^2)\beta_+ +l(l+2) \frac{\beta_- D_- }{D_+ }
\right)
\Biggr]  \; .
\end{equation}
For modes $l\geq 2$, the spherical droplet becomes unstable for $\bar R>R_l$ which in this limit is
given by 
\begin{equation}
R_{l} \approx \gamma  \frac{(4+3l+l^{2})D_+ \beta_+ +l(l+2)D_- \beta_- }{2D_+ (c_\infty -c^{(0)}_+ )}
\; .
\end{equation}
This expression shows that the the elongation mode $l=2$ is the first mode to become unstable. 
This provides an approximation for the critical radius of droplet division,
\begin{equation}
R_{\rm div}\simeq \gamma \frac{7\beta_+ +4\beta_- \frac{D_- }{D_+ }}{c_\infty -c^{(0)}_+ } \; 
\;,
\end{equation}
in the weak reaction limit. 
Because the limit  $R\ll  l_+$ corresponds to vanishing chemical reactions, this
approximate expression approaches the instability condition of the 
Mullins-Sekerka instability~\cite{Mullins1963}. 


\end{document}